\documentclass[9pt,twocolumn,twoside]{article}

\usepackage[utf8]{inputenc}
\usepackage{placeins}
 \usepackage{float}
 \usepackage{authblk}
 \usepackage{graphicx}

\title{Multi-channel Opto-mechanical Switch and Locking System for Wavemeters}

\author[1,*]{Moji Ghadimi}
\author[1]{Elizabeth M. Bridge}
\author[1]{Jordan Scarabel}
\author[1]{Steven Connell}
\author[1]{Kenji Shimizu}
\author[1,2]{Erik Streed}
\author[1,3]{Mirko Lobino}

\affil[1]{Centre for Quantum Dynamics, Griffith University, Nathan QLD 4111, Australia}
\affil[2]{Institute for Glycomics, Griffith University, Southport QLD 4215, Australia}
\affil[3]{Queensland Micro- and Nanotechnology Centre, Griffith University, Nathan QLD 4111, Australia}

\affil[*]{Corresponding author: Moji Ghadimi, m.ghadimi@griffith.edu.au}

\begin{document}

\maketitle

\begin{abstract}
\boldmath
Here we present a cost effective multi-channel opto-mechanical switch and software PID system for locking multiple lasers to a single channel commercial wavemeter. The switch is based on a rotating cylinder that selectively transmits one laser beam at a time to the wavemeter, the wavelength is read by the computer and an error signal is output to the lasers to correct wavelength drifts every millisecond. We use this system to stabilise 740\,nm (subsequently frequency doubled to 370\,nm), 399\,nm and 935\,nm lasers  for trapping and cooling different isotopes of Yb\textsuperscript{+} ion. We characterize the frequency stability of the three lasers by using a second, more precise, commercial wavemeter. We also characterise the absolute frequency stability of the 740\,nm laser using the fluorescence drift rate of a trapped \textsuperscript{174}Yb\textsuperscript{+} ion. For the 740\,nm laser we demonstrate an Allan deviation, $\frac{\Delta f}{f}$, of  $3 \times 10^{-10}$  (at 20\,s integration time), equivalent to sub-200 KHz stability. 
\unboldmath       
\end{abstract}

\section{Introduction}
Atomic physics covers a range of applications including quantum computing \cite{atom_qc_16, ion_qc_blatt_2008, ion_qc_blatt_2008_2, ion_qc_monroe_2013}, quantum simulation \cite{ion_qc_roos_2011, Safronova2017SearchMolecules}, quantum frequency metrology \cite{met_2015}, and tests of fundamental physics. The majority of these experiments rely heavily on laser systems, frequency stabilised to an atomic reference, for optical cooling \cite{b_cool_2007}, trapping and manipulating the electronic state of the atoms or ions being used as the quantum system. Traditionally the laser frequency would be locked to an atomic reference signal \cite{spec_2014, spec_1983}, or more recently, a stabilised optical reference cavity \cite{cav_2012, cav_2007}. The difficulty with these locking systems is the limited availability of atomic lines at the desired frequencies or the drift associated with optical reference cavities \cite{cav_2008}. Both approaches can only operate over a narrow design wavelength range. A relatively new alternative is to use wavemeters based on Fizeau interferometers \cite{WLM_2018, WLM_2015, WLM_2015_2, WLM_2012, WLM_2007}. These state of the art wavemeters can achieve sub MHz level precision making them suitable for stabilising the majority of laser systems in atomic physics experiments. They have the advantage that they are broadband, enabling the laser to be stabilised to any desired wavelength in their operating range (e.g. 400\,nm to 1100\,nm for the High Finesse WSU series), and with an appropriate optical switch they can become multi-channel, allowing the concurrent stabilisation of multiple laser systems at different wavelengths \cite{WLM_2018}. 

Similar to the optical reference cavities, the wavemeter is not inherently stable and is subject to drift, but due to the broadband nature and multi-channel capabilities they can be easily stabilised to a master laser system, stabilised to an absolute frequency reference,  operating anywhere within its wavelength range. This effectively transfers the stability of the master laser system to any desired wavelength within the operating range of the wavemeter. This is similar to the case of stabilising a mode-locked laser using frequency-domain stabilisation, in that the base frequency reference can be in the microwave or RF regime \cite{Telle1999Carrier-envelopeGeneration, Jones2000Carrier-envelopeSynthesis}. 

An optical switch is required to allow the concurrent stabilisation of multiple laser systems with one wavemeter. Commercial optical switches along with locking software are available to perform this task but they are costly, limiting their availability to many laboratories. 

Here we present a low-cost opto-mechanical 4-channel switch and 4-channel locking software that is used with a commercial WLM HighFinesse WSU-30 wavemeter to simultaneously lock multiple laser wavelengths. Unlike the commercial switch that can only be used with the expensive, top of the range, wavemeters, this system can work with any wavemeter that provides an accessible electrical or digital signal. It has no bandwidth limitation, 100\% transmission and perfect extinction. The design can readily be changed to accommodate more than four lasers by adding more openings to the cylinder and additional optics for overlapping the beams. 

We use lasers at wavelengths of 399\,nm, 740\,nm (frequency doubled to 370\,nm) and 935\,nm for photo-ionizing, cooling, and trapping of $^{174}$Yb$^+$ or $^{171}$Yb$^+$ ions for quantum physics experiments. We characterize the frequency stability of the three locked laser systems using a second higher resolution HighFinesse wavemeter. We also measure the absolute frequency fluctuations of the 740\,nm laser using the fluorescence fluctuations of $^{174}$Yb$^+$ ion as an absolute atomic reference. 

The mechanical switch does the job of alternating between the lasers with a duration of a few milliseconds (dependant on the rotation speed). Since the lasers are mechanically blocked, when one lasers is on, the others are perfectly extinguished. The 4-channel software interprets the different readings from the wavemeter and compares them with the desired frequency of each laser and uses digital to analog converter (DAC) devices to apply the necessary voltage to correct the frequency of that particular laser. The large difference between the wavelengths allows us to perform the separation of signals in software without the need for additional synchronisation information. All the material required to replicate the system including a 3D design of the cylinder and the LabVIEW software is available for download at https://github.com/Moji131/Wavemeter-Switch-PID-Labview. 

\section{Experimental Setup}
\begin{figure}[ht!]
\centering\includegraphics[width=0.5\textwidth]{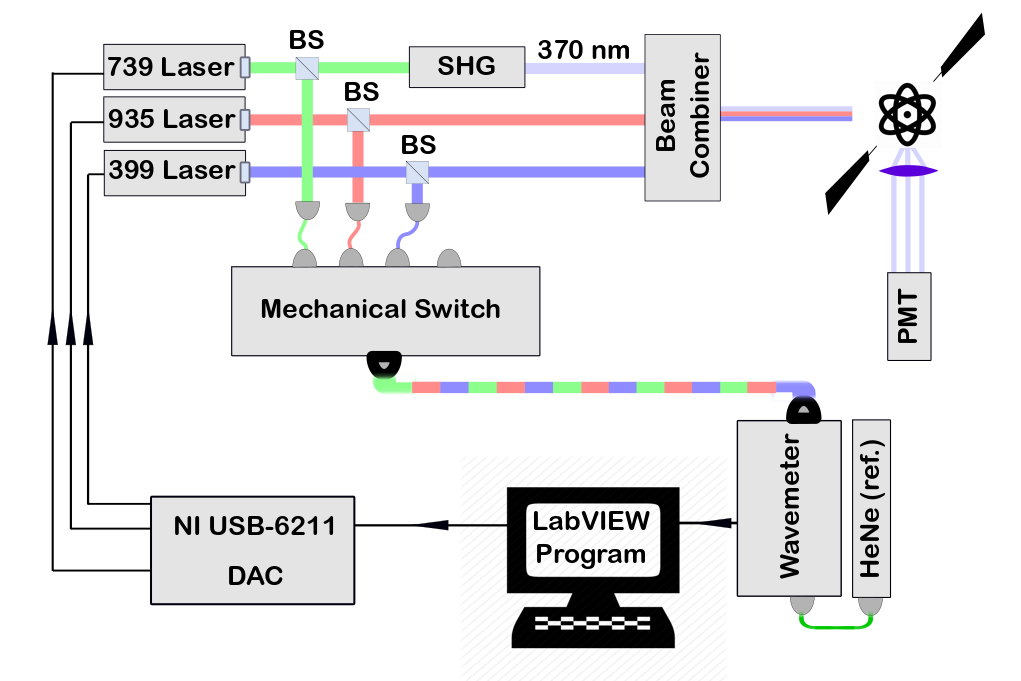}
\caption{ Schematic of the experiment for trapping and cooling a \textsuperscript{174}Yb\textsuperscript{+} or \textsuperscript{171}Yb\textsuperscript{+} ion (BS: Beam splitter, SHG: Second Harmonic Generation, PMT: Photo Multiplier Tube).  Three beam-splitters couple a fraction of the laser light into three single-mode optical fibres that go to the mechanical switch (see Fig. \ref{fig::switch} for details of the switch) which delivers the beams one at a time into a 830\,nm single-mode fibre that goes to the wavemeter. A LabVIEW program is used to to read the wavemeter output and ``lock'' the laser wavelengths. To mitigate long term drift of the wavemeter, it is automatically recalibrated to a reference HeNe laser every 10 seconds.}
\label{fig::exp}
\end{figure}

A schematic of our experimental setup is shown in Fig. \,\ref{fig::exp}. We use three lasers for trapping and cooling of a single \textsuperscript{174}Yb\textsuperscript{+} ion. Initially a 399\,nm laser is used for isotopically selective excitation of a neutral \textsuperscript{174}Yb atom from its ground state to $^2P_{1/2}$ state. From this state, we use a 370\,nm laser to ionize the atom \cite{Klein1990Laser+}. We use a M~Squared SolsTiS Ti-sapphire laser with wavelength of 740\,nm and a ECD-X second harmonic generation (SHG) box to generate this 370\,nm light. The same laser is then used for Doppler cooling of the ion. The S-P cooling transition of the ion is nearly closed, however with a probably of 0.5\% per cycle the ion may decay into the long-lived $^2D_{3/2}$ state. To prevent an interruption of the cooling process we use a 935\,nm laser to return the ion to~$^2S_{1/2}$ state via the $^3D[3/2]_{1/2}$ state \cite{Klein1990Laser+}.

We use a HighFinesse WSU-30 wavemeter and our software locking system to lock the wavelengths of the 740\,nm, 935\,nm and 399\,nm lasers. The wavemeter is recalibrated to a Thorlabs HRS015 stabilised HeNe laser ($\Delta \omega \approx$ 2\,MHz) every 10 seconds. In order to lock the lasers, a small portion of the power of each laser is coupled into single-mode optical fibers. These single-mode fibres take the light to a small breadboard which contains the mechanical switch and the optics shown in Fig. \ref{fig::switch} to overlap the beams such that they are all coupled into a single single-mode 830\,nm optical fibre. This fibre delivers the light to the wavemeter, as shown in Fig. \,\ref{fig::exp}. The mechanical switch ensures that only one laser is delivered to the wavemeter at a time, full details of the  switch are discussed in section \ref{sec::switch}.  The output of the wavemeter is read by a LabVIEW program that uses 4 separate software PID controllers to generate feedback signals, via a DAC, that go to the control electronics of the lasers. Details of the program are discussed in section \ref{sec::prog}. We were able to trap and cool either a \textsuperscript{174}Yb\textsuperscript{+} or a \textsuperscript{171}Yb\textsuperscript{+} ion using this wavemeter, switch and locking system and the stability of the lasers is discussed in section \ref{sec::res}.

\subsection{Mechanical Switch}
\label{sec::switch}
\begin{figure}[ht!]
\centering\includegraphics[width=0.5\textwidth]{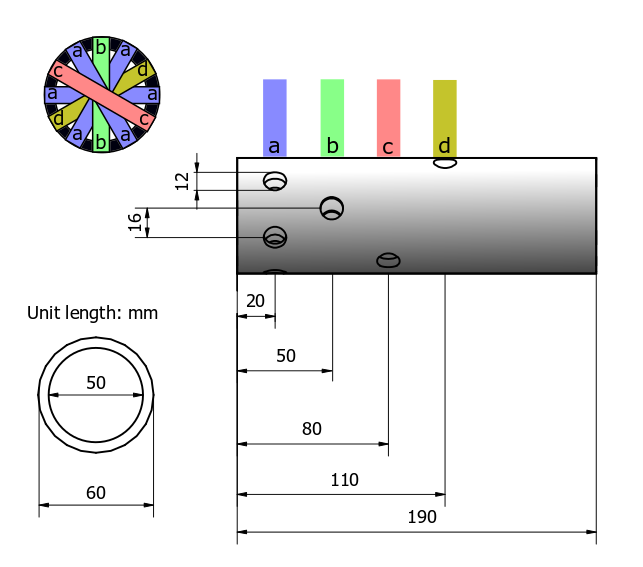}
\caption{Design of the cylinder used for the switching. Three pairs of holes on the left-hand end of the cylinder, allow the primary laser to pass six times a turn, whereas the other three lasers are allowed to pass twice per turn. The top
left image illustrates the duty cycle attributed to each laser, "a" sections represent the primary laser with three pairs of holes, and, "b", "c" and, 
"d" represent the secondary lasers with one pair of holes each. The black sections indicate where no lasers are passing. }
\label{fig::wheel}
\end{figure}

The main design considerations for the mechanical switch were that it needed to switch between multiple laser beams, letting only one beam through at a time, that it be easy to build and provide a cost-effective alternative to commercially available fibre switches. The main body of the switch is a rotating Delrin cylinder that has apertures positioned to allow four laser beams through, one at a time. The schematic of the cylinder is shown in Fig. \,\ref{fig::wheel}. The main cylinder is 60\,mm in diameter and 190\,mm in length. Six pairs of 12\,mm diameter holes run radially through the cylinder, positioned around its circumference with an azimuthal distance of 16\,mm between hole centres. Three of these pairs of holes are used for the primary laser input ("a" channels in Fig. \,\ref{fig::wheel}), the remaining three pairs are spaced along the length of the cylinder and are used for three other laser inputs. In our experiments the cooling laser has higher stability requirements than the other two lasers, so we chose to have multiple sets of holes for the primary input beam in order to allocate a higher proportion of the wavemeter read and feedback time to this laser system. 

A 24\,V UNITEMOTOR MY6812 Brush DC Motor is used to rotate the cylinder. By applying 15\,V to this motor we reach the speed of one revolution per 30\,ms (2000 RPM). At this speed the ``on''-time for each pair of holes is approximately 2\,ms, with a duty cycle of 37.5\% for the primary beam and 12.5\% for each of the other three beams. The dark time between two ``on''-times is  0.66\,ms. During each half-turn, the primary input beam will be delivered to the wavemeter three times and the others three beams are each delivered once. 

\begin{figure}[ht!]
\centering\includegraphics[width=0.5\textwidth]{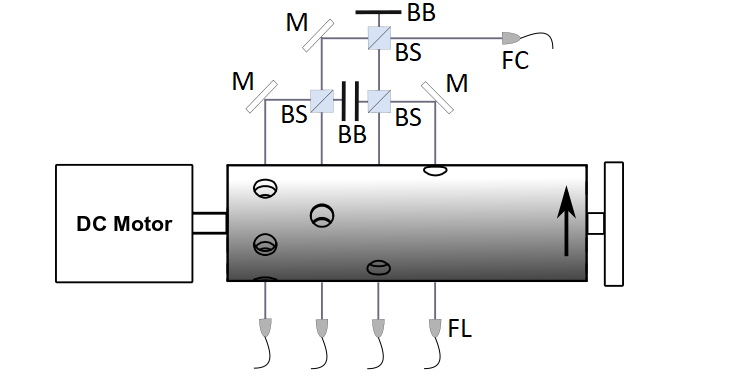}
\caption{Schematic of the mechanical switch and optical layout (M: mirror, BS: beam splitter, BB: beam block, FL: fibre launcher, FC: fibre coupler). The lasers are delivered using single-mode optical fibers specific to each laser. Rotation of the cylinder allows one laser to pass at a time. Lasers are overlapped using three broadband 50-50 beam splitter cubes and are coupled into a Thorlabs P1-830A-FC-2 fibre which is single-mode for 830\,nm to 980\,nm and few-mode for shorter wavelengths. This fibre delivers the alternating beams to the wavemeter. }
\label{fig::switch}
\end{figure}

Four single mode fibres are used to deliver the laser beams to the mechanical switch. In our setup we used Thorlabs P1-630A-FC-2 single-mode fibre for the 740\,nm laser, Thorlabs PM-S350-HP fibre for the 399\,nm laser and Thorlabs SM800-5.6-125 fibre for the 935\,nm laser. After the cylinder, three broadband 50-50 beam-splitter cubes (Thorlabs BS010) are used to overlap the beams, which are then coupled into a Thorlabs P1-830A-FC-2 fibre and delivered to the wavemeter. This last fibre is single-mode for 830\,nm to 980\,nm and few-mode for shorter wavelengths. Approximately $75\,\%$ of the power of each input beam is lost due to combining the beams using the 50-50 beam   splitters, a more efficient but less versatile approach would be to use carefully chosen dichroic mirrors to overlap the beams. A schematic of the switching system is shown in Fig.\,\ref{fig::switch}.

\begin{figure}[ht!]
\includegraphics[width=0.5\textwidth]{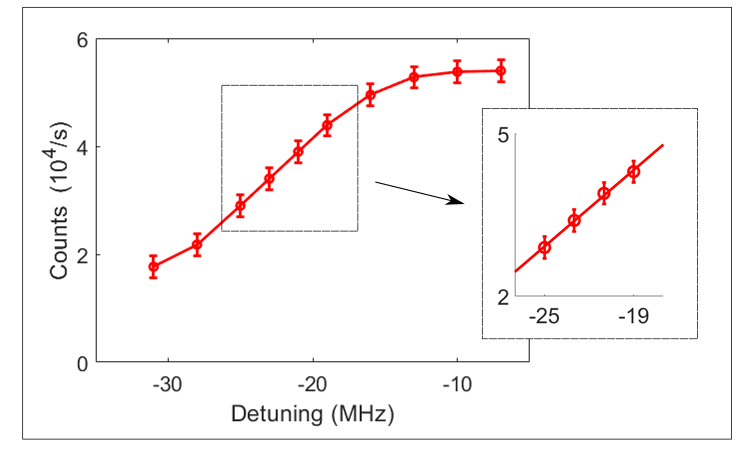}
\caption{Measured ion fluorescence counts vs. detuning of 370\,nm laser from resonance of the cooling transition of \textsuperscript{174}Yb\textsuperscript{+} ion. The curve is almost linear near -22\,MHz detuning. Inset:  A linear fit with MATLAB to the points near -22\,MHz shows a slope of $2.3 \times 10^3$ counts/s/MHz with 1-$\sigma$ error of 20 counts/s/MHz. This slope is used to convert fluorescence counts to frequency deviation of the 370\,nm laser and the result is halved to infer the frequency deviations of the 740\,nm laser.}
\label{fig::lw}
\end{figure}

\subsection{Locking Software}
\label{sec::prog}
In order to lock the lasers to the desired frequency we developed a LabVIEW program to interpret the wavemeter output and feedback to each of the laser controllers. The input to this program is the wavemeter output, which will either be the frequency of the incident laser or no reading if there is no incident laser at that time. On the LabVIEW front panel the user can specify a set frequency for the laser to lock to and an active frequency window for each of the four channels. If the reading from the wavemeter is within the active window of one of the channels, it will be treated as an input for that channel, compared to the set point and passed through the PID servo for that channel. The PID output value is passed to one of the National Instruments USB-6211 DACs, which generates the control voltage to feedback to the laser controller. For all the lasers cavity length is used as the tuning mechanism.  

When the wavemeter reading is outside the range of a particular channel, the channel remains temporarily inactive and the output voltage retains its current value until a new reading in that active window is available. The user must ensure that the active frequency windows do not overlap and that each initial laser frequency is inside the active locking window to enable the servo to start the locking process. Future improvements would include ordering to allow for multiplexing nearby wavelengths so long as they were not on neighbouring channels.

On the LabVIEW front panel each channel has a graph to display the deviation of the laser frequency in MHz from its set point, this assists the user in identifying if the laser is suitably locked and if the PID parameters are set appropriately. We have added the option for assigning two preset lock frequencies for all channels and buttons to easily toggle between them. We use this option to easily switch between the laser frequencies required for trapping two different isotopes of Yb (usually \textsuperscript{171}Yb\textsuperscript{+} and \textsuperscript{174}Yb\textsuperscript{+}).

\section{Results}
\label{sec::res}
\subsection{Absolute frequency instability measurements}
To characterise the absolute instability of the frequency of the 740\,nm laser, we monitored the fluctuations in the ion fluorescence rate during the steady state cooling process. The ion fluorescence rate is dependent on the detuning of the laser frequency from the resonance of the cooling transition. For \textsuperscript{174}Yb\textsuperscript{+} the natural full-width-half-maximum (FWHM) of the profile is 20\,MHz. In our experiment the FWHM is approximately 40\,MHz due to power broadening and the ion's micro-motion as a result of stray electric fields in the trapping region. Fig.\,\ref{fig::lw} shows the graph of ion fluorescence count rate vs laser detuning from the ion resonance close to the half-maximum point of the curve, the lineshape approximates a straight line, where the fluorescence rate is linearly proportional to the detuning. For the frequency instability measurements, the laser is locked to -22\,MHz detuning. We use an AOM to tune the detuning from -19\,MHz to -25\,MHz to calibrate the linear correlation between ion count rate and laser frequency. The inset in Fig.\,\ref{fig::lw} shows the linear fit. The linear fit has slope of $2.3\times 10^{3}$\,counts/s/MHz with $1-\sigma$ error of $\pm$\,20\,counts/s/MHz. 

We lock the laser frequency to this half-maximum value (detuning = -22\,MHz) so that changes in the ion fluorescence rate are linearly proportional to frequency deviations of the laser lock. We record the ion fluorescence for 500\,s, convert this to frequency fluctuations and calculate the Allan deviation \cite{Riley2008-NIST}:

\begin{equation}
\sigma(\tau ) = \sqrt{ \frac{1}{2(M-1)} \sum_{i=1}^{M-1}   ({\bar {y}}_{i+1}-{\bar {y}}_{i})^{2} } ,
\end{equation}

where ${y}_{i} = { \frac{\nu _{i}-\nu}{\nu} }$ is the $i_{th}$ of $M$ fractional frequency values averaged over the measurement (sampling) interval, $\tau$. Here $\nu_{i}$ is the i\textsubscript{th}  frequency measurement (averaged over the sampling interval) and $\nu$ is the nominal frequency.

The result of this measurement is shown in Fig.\,\ref{fig::740allan}.a. We repeat the measurement with the the mechanical switch disabled, so that the laser is continuously measured by the wavemeter, and show that in the same figure (\ref{fig::740allan}.a). We see the fractional frequency deviations are below $2 \times 10^{-9}$ at all measurement times for both scenarios, this corresponds to frequency deviations of 800\,kHz at the 740\,nm locking wavelength, which will be equivalent to 1.6\,MHz deviations after the frequency doubling process to the ultra-violet. After 1\,s of averaging time both data sets show frequency deviations of $<$200\,kHz at the locking wavelength. The plots show there is very little difference between the scenario where the wavemeter is constantly measuring the frequency (switching disabled) and that where it is measuring for 2\,ms in every 5\,ms (switching enabled). This suggests the switching is not adversely affecting the lock stability. It is important to note that during all the stability measurements in this paper, recalibration to the reference He-Ne laser was kept off. This introduces some drift that is not significant for a wavemeter in a box in a temperature-stabilised lab but it helps avoiding the injection of artificial noise during Allan deviation measurement as a result of random interruption of frequency reading at the time of recalibration.

\subsection{Relative frequency instability measurements}
Due to the lack of available closed cycling transitions for the ion at 399\,nm and 935\,nm we were not able to characterise the absolute frequency instability of theses lasers, instead we used a second wavemeter to charasterise their frequency instability. This second wavemeter (HighFineess Angstrom WS8-2) has a measurement resolution of 200\,kHz and accuracy of 2\,MHz. The locked laser to be measured is delivered to the WS8-2 via a single mode optical fibre. The fibre was connected through the commercial HighFineess switch but switching was disabled and only this particular port was being measured. The automatic recalibration to He-Ne laser was switched off during the measurements. 

\begin{figure*}[t]
\centering\includegraphics[width=1\textwidth]{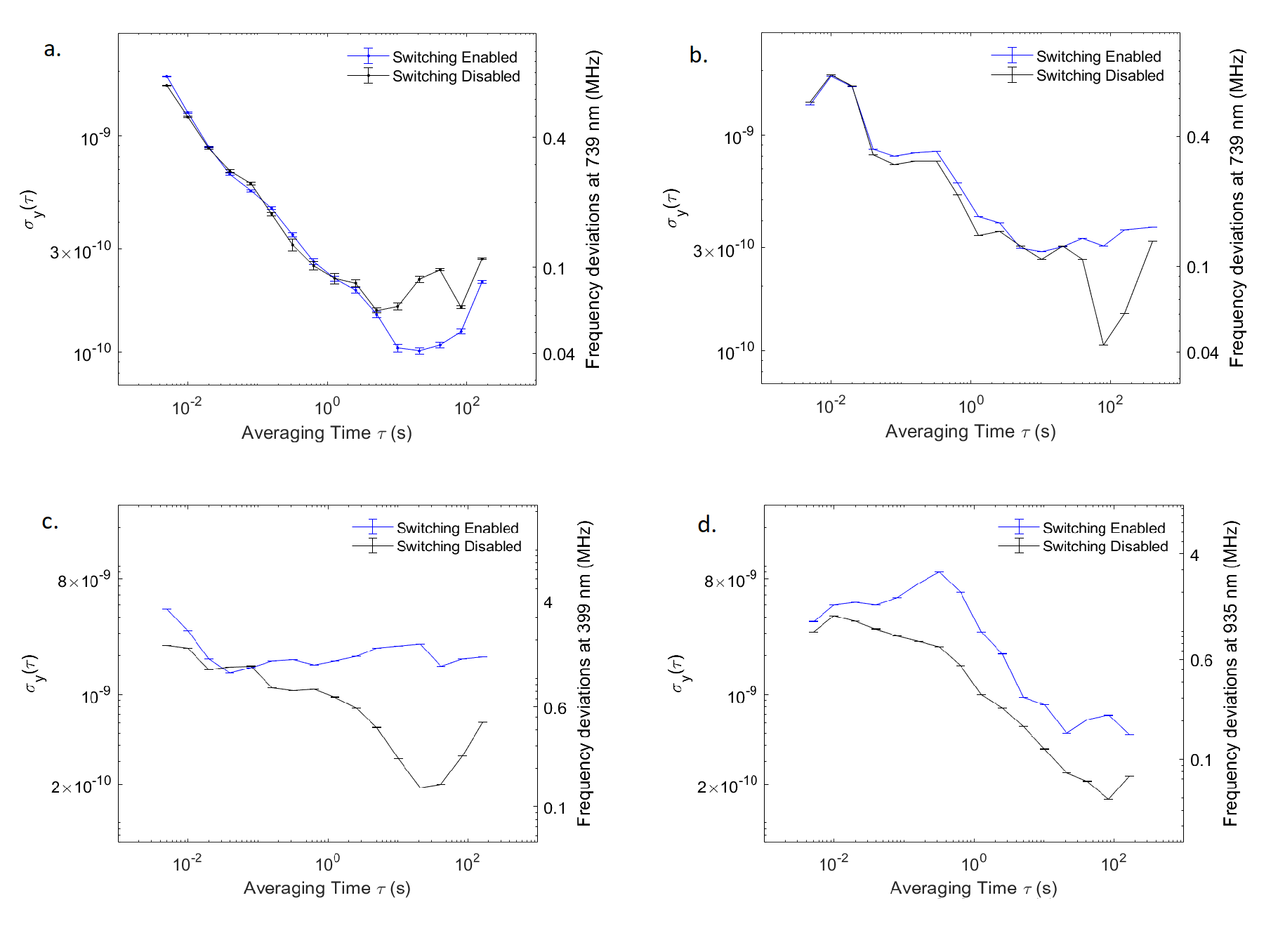}
\caption{Allan deviation measurements. a. 740~nm laser. The  frequency deviations are inferred from the fluctuations in the fluorescence count rate of the ion. Fractional frequency deviations remain below $2 \times 10^{-9}$ at all measurement times. There is little difference between the results when the switch is enabled or disabled.
b. 740~nm laser. The  frequency deviations are measured by a second wavemeter. The Allan deviation is broadly similar to a and shows little difference between the scenarios when the switch is enabled or disabled.
c. 399\,nm laser. When the switch is enabled, the instability worsens, but it is still within the the accuracy needed for our application.
d. 935\,nm laser. Similar to c, when the switch is enabled, the accuracy drops, but remains accurate enough for our experiments. 
The main difference between the locking of these lasers and the 740\,nm laser is that they are sampled only 2 times a cycle while the 740\,nm laser is sampled 6 times a cycle.}
\label{fig::740allan}
\end{figure*}

To test this measurement scheme, we started by making a measurement of the frequency instability of the locked 740\,nm relative to the second wavemeter. The resulting Allan Deviation is shown in Fig.\,\ref{fig::740allan}.b . We see the laser instability looks to follow the same trend as for the absolute fractional frequency instability measurement, it starts at $\sim 2 \times 10^{-9}$ and reduces to a minimum of a few $10^{-10}$ after 10 seconds. It also confirms that the instability does not differ significantly between having the switch enabled or disabled. This measurement is noisier than that using the ion, we attribute this to the relative drift of the two wavemeters, which are not recalibrated to the reference HeNe during the measurement and so are drifting relative to each other. The fact that the plots in Fig.\,\ref{fig::740allan}.a and Fig.\,\ref{fig::740allan}.b are broadly similar means that this is an appropriate way to characterise the frequency instability of the 399\,nm and 935\,nm lasers.

The results for characterising the frequency instability of the locked 399\,nm and 935\,nm lasers are shown in Fig.\,\ref{fig::740allan}.c and d respectively. We see that when the mechanical switch is disabled, and the wavemeter is continually measuring and feeding back to that laser, the fractional frequency instability follows the same trend as the 740\,nm laser. At short measurement times the instability is $\sim 2$ to $3 \times 10^{-9}$ and after 10\,s averaging time it reaches $\sim 2 \times 10^{-10}$. For both laser systems we see that enabling the mechanical switch, so that the wavemeter can lock multiple lasers, increases the instability of the locks. We observe that the 399\,nm laser lock fluctuates around a fractional frequency instability of $2 \times 10^{-9}$, which corresponds to frequency deviations of $\sim 1.5$\,MHz. The 935\,nm laser fractional frequency instability stays below a maximum of $9 \times 10^{-9}$, which is equivalent to 3\,MHz frequency deviations from the lock point. This stability is good enough for trapping Yb ions.
An important observation is that the lock instability of the 399\,nm and 935\,nm lasers increases when the mechanical switch is enabled, but the stability of the 740\,nm laser remains constant. The switch channel for the 740\,nm laser beam has extra sets of holes, which means the wavelength of this laser is measured three times as often as the other lasers, and this appears to significantly help with the lock performance for this particulat laser. If we needed better stability from the other lasers we could look at modifying the switch design to allow for more measurement time of these two lasers.  

\FloatBarrier

\section{Conclusion}
We have presented a multi-channel opto-mechanical switch for use with a commercial wavemeter to allow multiple lasers to be locked to one wavemeter. We developed a LabVIEW program to read the multiple laser frequencies from the wavemeter and feedback to the corresponding laser. We successfully used the system to lock three lasers to trap and cool a single \textsuperscript{174}Yb\textsuperscript{+} ion and a \textsuperscript{171}Yb\textsuperscript{+} ion. We characterised the frequency instability of the locked cooling laser using fluorescence of the ion and showed that it reaches a fractional frequency instability of below $10^{-10}$. The frequency instability of the other two lasers were characterised by a second wavemeter with higher frequency resolution and the fractional instability remained below $9 \times 10^{-9}$ for all measurement times. All the locked lasers remained within the natural linewidth of the electronic transitions we are driving in the ion. 

These results show the suitability of using a Fizeau wavelength meter and our mechanical switch for locking the lasers required to trap and cool ions or atoms in atomic physics experiments. This system can replace complicated and more costly atomic spectroscopy or Fabry-Perot cavity locking setups specific to each laser system. For absolute accuracy the wavemeter needs to be regularly calibrated to a stabilised laser, but this can be easily achieved with a commercial stabilised HeNe, which we find to be a simple and cost effective solution.

\section{Acknowledgment}
This research was financially supported by the Griffith University Research Infrastructure Programme, the Griffith Sciences equipment scheme, Australian Research Council Linkage project (LP180100096M), ML was supported by the Australian Research Council Future Fellowship (FT180100055), JS, SC, and KS were supported by the Australian Government Research Training Program Scholarship.


\bibliographystyle{unsrt}
\bibliography{bib1}

\end{document}